\documentclass[]{aa}
\usepackage{natbib}
\usepackage{graphicx}
\usepackage{epsfig,rotate}

\bibpunct{(}{)}{;}{a}{}{,}

\newcommand{\be}{\begin{equation}}
\newcommand{\ee}{\end{equation}}

\newcommand{\anrev}{{\it ARA\&A, }}

 \newcommand{\mnr}{{\it MNRAS, }}
\newcommand{\ana}{{\it A\&A, }}

\newcommand{\boldxi}{\mbox{\boldmath{$\xi$}}}

\begin{document}

\title{Protoplanet Magnetosphere Interactions}
\author{ John C. B.  Papaloizou }
 \authorrunning{  J.C.B. Papaloizou}
 \titlerunning{Protoplanet Magnetosphere Interactions}

\offprints{J.C.B. Papaloizou}

\institute{
Department of Applied Mathematics
and Theoretical Physics, University of Cambridge, Centre for
Mathematical Sciences, Wilberforce Road, Cambridge, CB3 0WA, UK \\ \email{jcbp2@damtp.cam.ac.uk}}

\date{Accepted; Received; in original form;}

\label{firstpage}

\abstract
{}
{In this paper, we study a simple model of an orbiting protoplanet
in a central magnetospheric cavity, the entry into such a cavity having been proposed
as a mechanism for halting inward orbital migration.
  }
{We  have calculated the gravitational interaction 
 of the protoplanet with the magnetosphere 
 using a local  model
and determined the rate of   evolution of the orbit.
  }
{ The interaction is found to be determined by the outward flux
of MHD waves and thus the possibility of the  existence of such waves in the cavity
is significant.
  }
{The estimated orbital evolution rates due to gravitational and other interactions
with the magnetosphere  are unlikely
to be significant during protoplanetary disk lifetimes.
  }


 \maketitle

\section{Introduction}\label{sec1}
The discovery of  extrasolar giant planets
orbiting close to their host stars ( Mayor \& Queloz 1995; Marcy \&
Butler 1995, 1998) with periods of a few days has led to the proposal that orbital migration
during and post formation is responsible for them attaining short period
orbits.
This is because of the  difficulties associated with forming such planets in situ, either in the critical 
 core mass accumulation followed by gas accretion scenario, or the gravitational 
 instability scenario for giant planet formation (see Papaloizou \& Terquem 2006 and references therein
 for more discussion).
 However, estimates of disk driven migration time scales have in general found these to be
 short compared to protostellar disk lifetimes (e.g.  Nelson et al. 2000). 
 This in turn suggests the need for a mechanism to halt migration so preventing the
 protoplanet from falling into the star. Lin Bodenheimer \& Richardson (1996)
 suggested that entry into a stellar magnetosphere close to the star could provide
 such a mechanism. This is through detachment from the disk. Indeed once the protoplanet is       interior to the 2:1 resonance with the inner disk edge, interaction with the disk would be       expected to cease through the lack of effective outer Lindlbad resonances 
 (see e.g. Lin \& Papaloizou 1993). 

 However, 
 there has as yet been no  detailed analysis of the expected orbital evolution
 of protoplanets  in magnetospheric cavities or demonstration of the effectiveness
 of this mechanism for halting inward orbital migration. This would require
 consideration of models for this inner region and a  calculation of the
 interaction with the protoplanet and the effects on the protoplanet orbit.
 It is the purpose  of this paper to make some first steps in this direction.

Accretion on to T Tauri stars  is indeed thought  to be magnetically
dominated close to the central star in some cases (Cameron \& Campbell 1993; Menard et al. 2003).
 Surface field strengths
of a few kilogauss (e.g. Safier 1998; Johns-Krull et al. 1999) have been estimated.
According to  models of accretion flows with their central regions
dominated by the magnetic field
of the central star ( e.g. Ghosh \& Lamb 1978; Cameron \& Campbell 1993;
Matt \& Pudritz 2004) , at large radii the accretion flow
is that of a normal viscous accretion disk ( e.g. Pringle 1981). 
The magnetic field
penetrates the disk due to diffusion arising
from the growth of various instabilities.  At large radii the magnetic field is weak
such that viscous stresses are more important than magnetic stresses. 
Then the effect of the magnetic field on the disk structure can be ignored.
However, at smaller radii the dominance of magnetic stresses 
leads to the truncation of the  disk at a radius,
$R_{\rm d,}$ and the channeling of the flow to the star along
stellar magnetic field lines. 

In order for  accretion  on to the central star to
proceed the magnetic field must not
disrupt the disk  exterior to the corotation radius, where
the angular velocity in the disk coincides with that of the star,  since the
magnetic torque  on the disk would
impart  angular momentum to 
the disk gas,   making it unable to connect to the stellar field.
If  disk disruption  occurs too far inside corotation the  radius,
there is  an  accretion torque acting  to spin up the star.
In general one  can expect  there to be an equilibrium state 
where spin up torques due to accretion  and spin down torques due
to magnetic stresses acting on the disk cancel.
Because of  the expected rapid increase of the magnetic stresses
as the distance to the star decreases, one expects that
$R_{\rm d}$ is slightly less than the corotation radius
(Ghosh \& Lamb 1978) when the system accretes onto the central star
and is near equilibrium. 

 This general picture of magnetic accretion has been found in  recent three dimensional
numerical simulations   (see  Romanova et al. 2004; 2006; Bouvier et al. 2006).
It has been found that accretion along field lines takes place in funnel stream  flows.  When the
stellar field is a dipole that is either  not nearly aligned or
  strongly misaligned with the angular momentum axis,
the magnetospheric density is relatively low in the equatorial plane.
 For both small and  large misalignment angles,
 relatively high density streams may occur there. 
 The latter situation may be less favourable for stalling migration as 
interaction with the protoplanet is potentially stronger.

The purpose of this paper is to consider a simple model of the stellar magnetosphere
with a steady state accretion flow and to use it to calculate the protoplanet magnetosphere
interaction and the consequent protoplanet orbital evolution rate. This is to
establish to what extent entry into such a magnetosphere is by itself sufficient to halt migration
without the need to postulate additional effects such as residual coorbital
torques resulting from the inner disk edge ( e.g. Masset et al. 2006).

The plan of the paper is as follows:
In section \ref{sec2} we give the basic equations of the problem.
In section \ref{sec4} we consider perturbation of  a steady state magnetosphere
by  a protoplanet in circular orbit  with the aim of
evaluating the exchange of energy with the orbit resulting from
the perturbed gravitational forces.

We carry out a purely local calculation of the protoplanet magnetosphere interaction.
This confirms the importance of the existence of propagating waves in order
to obtain a non zero energy exchange rate. When the flow is highly super sonic and super Alfv{\'e}nic the interaction can be described  using the well known  dynamical friction
formalism of Chandrasekhar (1943) and  Tremaine \& Weinberg (1984).
However, when the flow speed is below that of the fast magnetosonic speed but
significantly  exceeds the slow magnetosonic speed, as in our case, the energy exchange rate is 
found to be reduced.
This is particularly the case when the relative velocity between gas and protoplanet
is parallel or nearly parallel to the magnetic field. A similar reduction  would be expected
to apply to the accretion rate by the protoplanet.

Using the conservation of energy applied to the fluid perturbations
 in a more general context,
 we find that the energy exchange rate with the orbit is determined by
advected and wave energy fluxes at the system boundaries, confirming the importance of the existence
of propagating waves and the boundary conditions for the determination of the energy exchange rate with the orbit.

In section \ref{sec5} we discuss our results, showing that, for parameters
expected for protoplanetary disks,  these
imply negligible orbital evolution for protoplanets  interior
to magnetospheric cavities. This in turn implies that if  entry
into a magnetospheric cavity is responsible for a lack of further migration
of a single protoplanet,
 the orbital elements could subsequently only be affected by stellar tides.
 Finally we  summarise  our conclusions.

\section{Basic Equations}\label{sec2}
The basic equations are the equations  of ideal MHD in a frame rotating
with uniform angular velocity ${{\bf e_3}}\Omega_{*},$
where, adopting cylindrical polar coordinates $(r ,\varphi, z)$ with origin at the central
star,  ${{\bf e_3}}$ denotes the unit vector in the  $z$ direction
and $\Omega_*$ is the angular velocity of the central magnetospheric region,
taken to coincide with that of the central star.

 The equation of motion can be written in the form 
\be {\partial {\bf v }\over \partial t}+ {\bf v}\cdot \nabla{\bf v}
+2\Omega_*{ {\bf e_3}}\times {\bf v}
= {\bf f} \label{mot},\ee where
the force per unit mass is given by
\be {\bf f} =
-{{\nabla P}\over \rho }-\nabla \Phi+ {(\nabla \times {\bf B})\times {\bf B}\over 4\pi \rho}, \ee 
the induction equation in the form 
\be {\partial {\bf B}\over \partial t} = \nabla \times ({\bf v}\times {\bf B})\label{induct} \ee
and the continuity equation in the form 
\be {\partial \rho \over \partial t} + \nabla\cdot(\rho {\bf v}) =0 .\label{cont}\ee
Here $\rho$ is the density, $P$ is the pressure, ${\bf v}$ is the velocity and the magnetic field is ${\bf B}.$

\noindent When  the only source of gravity is
the central star treated as a point mass, the combined gravitational and centrifugal potential is
\be \Phi = -{GM_*\over \sqrt{r^2+z^2}}- {1\over 2}\Omega_*^2r^2 ,\ee
with the central stellar mass being $M_*$ and $G$ the gravitational constant.
For the model considered here we adopt an isothermal equation of state
so that $P  = \rho c^2$ with $c$ being the constant sound speed.
We take this to be characteristic of the outer disk so that it
will be much less than the orbital speed or a characteristic Alfv{\'e}n speed.

\section{Perturbation Due to an Orbiting Protoplanet}\label{sec4}
\subsection{Angular Momentum Transport and Equilibrium Stellar Spin}
 We consider the perturbation of a magnetosphere interior to an  accretion disk through which
matter accretes at a constant or slowly varying rate by an orbiting protoplanet.
When a quasi-steady state has been reached, the disk is expected to be disrupted 
at a radius, $R_d$ slightly interior to the corotation radius, $r_0$
where the stellar angular velocity coincides with the Keplerian angular velocity
expected in the disk.
This is given by 
\be r_0 =(GM_*/\Omega_*^2)^{1/3}.\ee
In the inner region the flow is expected to become magnetically dominated and sub Alfv{'e}nic
(Ghosh \& Lamb 1978). This general behaviour with
the magnetospheric radius being $\sim 0.85r_0$ has  been
found in recent simulations (see  Bouvier et al. 2006).
It has also been estimated that the time required to attain such
a quasi steady state is short compared to the lifetime of the  disk
  (e.g. Cameron \& Campbell 1993; 
Armitage \& Clarke 1996). For a steady state flow (\ref{induct}) and (\ref{cont})
imply that ${\bf v}$ and ${\bf B}$ are parallel.

\subsection{Local Analysis}
Finding the response of a general steady state magnetospheric
accretion flow to an orbiting protoplanet is a very difficult problem.
Accordingly we consider possible simplifications.
In this section we consider the calculation of the response using a local approximation.
The idea behind this is that at any stage, the protoplanet moves relative to field lines
along which the accretion flow moves, such that the speed of the
 protoplanet relative to the gas, denoted by $|{\bf U} - {\bf v}|$ below, may become comparable to the orbital speed (see Figure~1). The characteristic scale of interaction  at which
a protoplanet of mass, $M_p,$  would dominate the  gas flow would then be set by the classical
Bondi-Hoyle radius (e.g. Hoyle \& Lyttleton 1939; Ruffert 1996) $r_{BH} = 2GM_p/|{\bf U} - {\bf v}|^2.$
When the relative speed is characteristically orbital, $r_{BH}$ 
is smaller by a factor  $ \sim M_p/M_* \ll 1$
than the  radius at which the  protoplanet orbits indicating a local interaction
on a characteristic time scale that is smaller than the orbital time by a similar factor.
The smallness of $r_{BH}/r_H \sim (M_p/M_*)^{2/3},$ with $r_H$  being the Hill radius
also indicates that the interaction of the protoplanet with gas that is not bound to it
is linear even in the giant planet regime with $M_p/M_* \sim 10^{-3}.$
The margins involved are such that the interaction is expected to be local and linear even  when the relative speed
is significantly less than orbital.

This is unlike the situation in protoplanetary disks where the small relative 
velocity between gas and protoplanet can make the interaction nonlinear at scales
larger than the Hill radius resulting in gap formation (see e.g. Lin \& Papaloizou 1993; 
 Papaloizou \& Terquem 2006; and references therein  for more discussion).

\begin{figure*}
\centerline{\epsfig{file=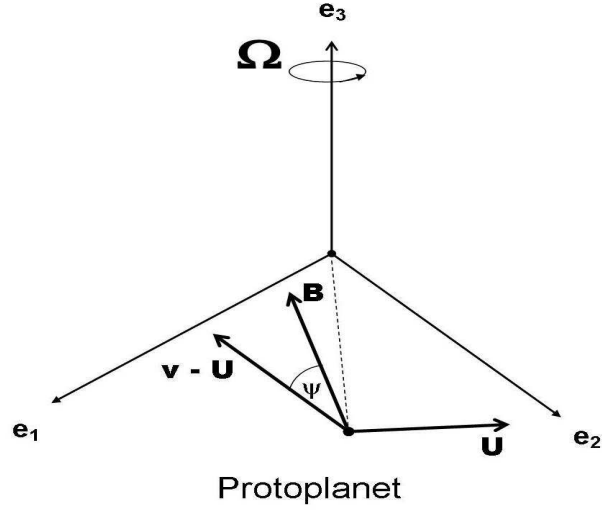 ,width=10cm, height = 8cm, angle=0}}
\caption{\label{fig1} This illustration shows the protoplanet viewed
in a frame rotating with the angular velocity of the star $\Omega_*$ here $\equiv \Omega,$
the latter being located at the origin of a Cartesian coordinate system
$({\bf e_1}, {\bf e_2},{\bf e_3})$ defining the direction of the axes. The protoplanet moves in a circular orbit in the equatorial plane with velocity ${\bf U}.$ 
The field line passing through its location points out of the plane
ultimately reaching the star. In this frame the gas moves along field 
lines with velocity ${\bf v}$ and the line pointing in the
 direction of the relative velocity ${\bf v-U}$
makes an angle $\psi$ with that pointing in the direction of  ${\bf B},$ 
or equivalently ${\bf V_A},$ which also gives the direction of the $z$ axis for the local frame
in which the calculation is done.
In that frame ${\bf v-U}$ lies in the $(x,z)$ plane.}
\end{figure*}

On the basis of the above discussion, we here adopt the assumption that the protoplanet magnetosphere interaction
is localised in the vicinity of the protoplanet and that the associated characteristic time
scale is significantly shorter  than  the orbital period.
Accordingly  we  adopt a local frame that can be taken to be a
non rotating  local Cartesian coordinate system with origin instantaneously at the
centre of mass of the protoplanet and moving with a constant velocity equal
to that of the rotational velocity the star would have if it extended to  the radial location of the protoplanet
(if one uses a local frame fixed in the original rotating frame
and then neglects $\Omega_*$ in the equations of motion the same results are obtained).
 The response to the protoplanet is readily calculable  
and leads to  a confirmation of the conclusion  that the changes induced in the orbit are associated with
 wave energy fluxes propagating away from the protoplanet. Furthermore when such
waves are absent, such as when the relative speed is below all possible wave speeds,
there is no energy change induced on the orbit.

\subsection{ Linear Analysis}

The effect of a perturbing protoplanet of mass $M_p$ at position vector 
${\bf r}={\bf r}_{orb}$ is to produce an Eulerian perturbation to the gravitational potential
 $\Phi' = -GM_p/\sqrt{|{\bf r} -{\bf r}_{orb}|^2+b^2},$
 where $b$ is a softening parameter which  can be specified to ensure 
 $\Phi'$ remains finite. An appropriate value
 is expected to be of the  order of the radius of the planet. 
  
In the original rotating frame, the Lagrangian displacement  ${\mbox{\boldmath$\xi$}}$
induced by by this perturbing potential
satisfies the equation (see eg. Frieman \& Rotenberg 1960;
Lynden-Bell \& Ostriker 1967)

\be {D^2 \boldxi \over Dt^2} + 2\Omega_* {\hat {\bf e_3}}\times {D \boldxi \over Dt}
= F(\boldxi),\label{Lagp}\ee
where the convective derivative operator
\be {D \over Dt} \equiv {\partial \over \partial t} + {\bf v}\cdot\nabla  \ee
and  the Lagrangian variation
of the force per unit mass is
$F(\boldxi) = {\bf f}' + \boldxi\cdot \nabla {\bf f},$ with
$$ {\bf f'} =
-{{\nabla P'}\over \rho } +{\rho'\over \rho^2 }\left(\nabla P -{(\nabla \times {\bf B})\times                          {\bf B}\over 4\pi}\right) \hspace{10mm}$$
\be\hspace{1.5cm} +{(\nabla \times {\bf B'})\times {\bf B}\over 4\pi \rho}
+{(\nabla \times {\bf B})\times {\bf B'}\over 4\pi \rho} -\nabla \Phi'. \ee
 
\noindent The Eulerian perturbations to the magnetic field  and density are given by
\be {\bf B}' = \nabla \times (\boldxi \times {\bf B})\label{PF} ,\ee
and
\be \rho' = -\nabla\cdot (\rho \boldxi) \label{indp},\ee
respectively.
For an isothermal equation of state the Eulerian pressure perturbation $P'=\rho' c^2.$
Equations (\ref{Lagp} - \ref{indp}) thus provide a complete system
for determining the Lagrangian displacement $\boldxi.$

We  make the usual local
approximation that the  steady state  variables $\rho,$ $\Phi$ and $P$ as well as ${\bf v}$
and ${\bf B}$ can be taken to be  constant.
In the local  frame ( see Figure~1) the velocity of the protoplanet is taken to be ${\bf U}$
and we may use equations (\ref{Lagp} - \ref{indp}) with $\Omega_*=0.$. 
 Equation (\ref{Lagp}) then gives
\be \left ({\partial \over \partial t}+ {\bf v}\cdot \nabla \right)^2 \boldxi 
= -{{\nabla P'}\over \rho }-\nabla \Phi'  
 +{(\nabla \times {\bf B'})\times {\bf B}\over 4\pi \rho} 
,\label{Lagploc}\ee
which together with equations (\ref{PF}), (\ref{indp}) and $P' = -\rho' c^2$ 
enable the Lagrangian displacement induced by  the perturbing
potential $\Phi'$ to be calculated.

\subsubsection{Fourier Transform of the Potential}
For this local model, we set the softening parameter $b=0,$ so
that in the local coordinates, the perturbing potential may be written
$\Phi' = -GM_p/|{\bf r} - {\bf U}t|,$  with
${\bf r}$ denoting the position vector. To solve (\ref{Lagploc}) we take  Fourier transforms
of the perturbations. Thus for a general perturbation quantity $Q,$
\be Q= \int \exp(i{\bf k}\cdot {\bf r})Q_{\bf k}({\bf k})  d^3 {\bf k},\ee  with the Fourier
transform
 \be Q_{\bf k} ={1\over (2\pi)^3} \int \exp(-i{\bf k}\cdot {\bf r})Q ({\bf r}) d^3 {\bf r}.\ee
 The wavenumber ${\bf k} = (k_x, k_y, k_z)$ and the integrals are taken over
 all coordinate and wavenumber space.
 For the perturbing  potential $\Phi',$ one readily finds that
 \be \Phi'_{\bf k} = -{ GM_p\over 2\pi^2|{\bf k}|^2} \exp(-i{\bf k}\cdot {\bf U}t). \ee 
\subsubsection{Calculation of the Fourier Transform of the Displacement}
After taking the Fourier transforms of equation (\ref{Lagploc})
and  equation (\ref{PF}) and noting that 
$P'_{\bf k} = \rho'_{\bf k}c^2 =  -\rho c^2 \boldxi_{\bf k}
\cdot {\bf k},$ one obtains a set of linear algebraic equations
for the determination of the components of $\boldxi_{\bf k}$ in terms of 
$\Phi'_{\bf k}.$  The scalar quantity $\boldxi_{\bf k}
\cdot {\bf k},$ is readily found after straightforward algebra to be given by
\be \boldxi_{\bf k}\cdot {\bf k}= {i\Phi'_{\bf k}{\bf k}^2( \nu^2
-  ({\bf k\cdot V_A})^2)\over (\nu^4 -(c^2+V_A^2)\nu^2
{\bf k}^2+ ({\bf k\cdot V_A})^2c^2{\bf k}^2)}, \label{locresp}\ee
where $\nu=({\bf k\cdot (v-U)})$ and
${\bf V_A} =  {\bf  B}/\sqrt{4\pi \rho}$ denotes a vector, with magnitude 
equal to the Alfv{\`e}n speed, and the direction of ${\bf B}.$ We set $V_A= |{\bf V_A}|.$

The gravitational force acting on the protoplanet as a result of
the perturbation of the magnetosphere  can then be found using
\be {\bf F_{orb}} =  \int \rho' \nabla \Phi' d^3{\bf r} .\label{FO}\ee
We evaluate the rate of change of orbital energy that would be viewed in the fluid
rest frame. In that frame the drag force would be zero if ${\bf v}= {\bf U}$
and consequently the rate of change of orbital energy would be zero. Otherwise it is given by
\be {dE_{orb}\over dt} = {\bf (U-v)\cdot F_{orb}} \ \ {\rm or} \ee
\be {dE_{orb}\over dt} = 
-(2\pi)^3 Re\int \rho  (\boldxi_{\bf k}\cdot {\bf k})({\bf k \cdot( U-v)})\Phi'^{*}_{\bf k} d^3{\bf k} .\label{force}\ee
Here $Re$ indicates that the real part of the integral is to be taken.

In evaluating equation (\ref{force}) we note that if there were no
singularities in the expression for $\boldxi_{\bf k}\cdot {\bf k}$
 given by equation (\ref{locresp}) we would have no energy exchange
 or $ {dE_{orb}/dt} =0.$ Thus in  order to obtain
 $ {dE_{orb}/dt} \ne 0,$ there should be singularities in $\boldxi_{\bf k}\cdot {\bf k}.$   
 In fact singularities  in  the form of simple
 poles occur when the denominator on the right hand side of
 equation (\ref{locresp}) vanishes or when
 \be \nu^4 -(c^2+V_A^2)\nu^2
{\bf k}^2+ ({\bf k\cdot V_A})^2c^2{\bf k}^2 = 0.\label{FWD}\ee
This quartic gives  four roots for $\nu$ given by
$\nu = \pm r_{1,2}|{\bf k}|,$
where $r_1^2 = (c^2+V_A^2)/2 + \sqrt{(c^2+V_A^2)^2/4 - ({\bf{\hat  k}\cdot V_A})^2c^2}$
and  $r_2^2 = (c^2+V_A^2)/2 - \sqrt{(c^2+V_A^2)^2/4 - ({\bf{\hat  k}\cdot V_A})^2c^2},$
with ${\bf {\hat k}} = {\bf k}/|{\bf k}|.$

The roots corresponding to $r_1$ and $r_2,$ which without loss of generality
are taken to be positive,  correspond to fast and slow magnetosonic waves
respectively which propagate with angular frequency $\nu ={\bf k\cdot (v-U)}.$ 
When any such waves can be present, there is the possibility of
an induced change of orbital energy. 
This is entirely consistent with the fact, noted in section \ref{WA} below that the rate of change of orbital
energy has to be associated with a conserved energy flux propagating 
away from the protoplanet. In the local problem considered here,
this flux can be carried in either fast or slow magnetosonic waves.
We also comment that in order to produce orbital energy changes, 
as is apparent from equation (\ref{FO}),
the waves have to be compressive or associated with density perturbations,
thus incompressible Alfv{\'e}n waves are ineffective..

The singularities in $\boldxi_{\bf k}\cdot {\bf k}$
given by (\ref{locresp}) can be dealt with by writing the denominator 
as a product of factors of the form $({\tilde \nu} \pm r_i),$ for  $(i= 1,2)$
${\tilde \nu} = \nu/|{\bf k}|$ and then handling the inverse of
each these by using the well known Landau prescription, thus
\be {1\over ({\tilde \nu} \pm r_i)} \rightarrow P \left ({1\over ({\tilde \nu} \pm r_i)}\right)
+\pi i\delta({\tilde \nu} \pm r_i). \ee
Here $P$ denotes that the principal value is to be taken on integration
and $\delta$ is Dirac's $\delta$ function. This prescription results from adding
an infinitesimally small negative imaginary part to ${\tilde \nu},$ which can be regarded
as ensuring that the perturbing potential vanishes at $ t = -\infty$ thus
imposing causality. Using the Landau prescription  and noting that roots of opposite
sign give equivalent contributions, we can find the rate
of change of orbital energy using equation (\ref{force}) in the form
\be  {dE_{orb}\over dt} = -8(\pi)^4\rho\int {|\Phi'_{\bf k}|^2 {\tilde \nu}|{\bf k}|\over r_1r_2(r_1^2-r_2^2)}\Lambda({\bf {\hat k}}) d^3{\bf k},\ee where
$$ \Lambda =  r_2(r_1^2 -({\bf {\hat k}\cdot V_A})^2)
\delta({\tilde \nu} -r_1)$$
\be \hspace{2.5cm} -r_1(r_2^2 -({\bf {\hat k}\cdot V_A})^2)
\delta({\tilde \nu} -r_2)
\ee
To perform the above integral we first remark that because $r_1,r_2$ and ${\tilde \nu}$
are functions only of the unit wavenumber vector, if polar coordinates in wavenumber space
are adopted, the integrand easily factors into radial and angular parts, thus
\be  {dE_{orb}\over dt} = -2(GM_p)^2\rho\int_{k_{min}}^{k_{max}}{dk\over k}
\int {{\tilde \nu \Lambda({\bf {\hat k}})}\over r_1r_2(r_1^2-r_2^2)} d\Omega, \label{aint}\ee where
$d\Omega$ denotes integration over the solid angle in wavenumber space
and $k_{max}, k_{min}$ are the magnitudes of upper and lower wavenumber cut offs
that necessarily have to be used in local calculations of gravitational
interactions because of the 
long and short range form of the $1/r$ potential (see eg. Tremaine \& Weinberg 1984 in the context of
stellar dynamics and 
Papaloizou 2002 in the context of thin hydrodynamic disks).
Physically $k_{max}^{-1}$ should be taken as the smallest effective
length scale in the problem and $k_{min}^{-1}$  the largest.
 For a giant protoplanet in an inner magnetosphere, the ratio of the radius of the planet
to a characteristic scale of the  magnetosphere of $0.1AU,$ $\sim 10^{-2},$  accordingly one expects
$k_{max}/ k_{min}\sim 100.$

To perform the angular integral in equation(\ref{aint}), we specify the local frame
such that the $z$ axis points along the direction of ${\bf B}$ and thus ${\bf V_{A}}$
and ${\bf v}.$ The direction of ${\bf v -U}$ is taken to be in the $(x,z)$ plane
making an angle $\psi$ with the $z$ axis (see Figure~1). Using spherical polar angles $\theta, \phi$
to perform the integral over the solid angle, we note that   because ${\bf{\hat  k}\cdot V_A}/V_A = \cos\theta,$
 $r_1,r_2$ are functions
only of $\theta.$   Given that ${\tilde \nu}~=~|{\bf v-U}|~(\cos\theta\cos\psi~+~\sin\theta\sin\psi\cos\phi~)~,$ the delta functions
readily enable the integration over $\phi$ and so we obtain an expression requiring only
an integration over $\mu = \cos \theta$ in the form
\be  {dE_{orb}\over dt} = -{4\pi(GM_p)^2\rho\over |{\bf v-U}|} \ln\left ({k_{max}\over k_{min}}\right){\cal D}, \label{bint}\ee
 where $ {\cal D} = (I_{1,2}+I_{2,1})/\pi,$ with
\be I_{1,2} =  \int_{D}
 {(r_1^2 - V_A^2\mu^2)d\mu \over (r_1^2-r_2^2)\sqrt{(1-f_1^2)\sin^2\psi -(\mu - f_1\cos\psi)^2}}
 , \label{I12}\ee
 where $(f_1 , f_2)= (r_1 , r_2) /|{\bf v-U}|.$
 In the above integral the domain of integration is either $(-1,1)$ or
 $(s_1,s_2),$ where $s_1$ and $s_2$ are values of $\mu$ for which the square root
 in (\ref{I12}) is zero should they exist.
 
 In general the dependence of $f_1$ and $f_2$ on $\mu$ through $r_1$ and $r_2$
 makes the above integral intractable. However, it can be found
 in the limit of low $c/V_A$ that we consider here. Then $r_1 = V_A,$ and
 $r_2 = c\mu.$
 
 We first note that 
  In the  super Alfv{\'e}nic case where both $f_1$ and $f_2$ are very small, one
 obtains ${\cal D}  = 1.$ This case of course applies to the limit of zero wave propagation
 speeds which corresponds to the dynamical friction
 calculation of Chandrasekhar (1943)  and it can also  be derived from the calculation
 of the  linear response
  of collisionless particles
  by Tremaine \& Weinberg (1984). We obtain ${\cal D} \le 1$
  so that then this quantity measures the factor  by  which the dynamical friction is
  reduced when compared to the estimate of Chandrasekhar (1943).
  
  To emphasise that this can be a large reduction, we note that in
   the  sub Alfv{\'e}nic case when both $f_1$ and $f_2$ exceed unity, there is
 no wavelike response or domain of integration for $(I_{1,2} , I_{2,1})$ so that
  ${\cal D} = 0.$
 
 When $f_1$ is large but $f_2$ very small so that slow waves exist but fast waves do not,
 we have  $I_{1,2} =0 $ and $I_{2,1} = (\pi \sin^2\psi)/2$ so that ${\cal D} =(\sin^2\psi)/2 .$
 
 \noindent  Thus in this case the rate of change of orbital energy depends on the orientation of the
 magnetic field, being zero when $\psi =0.$  In this limit ${\bf U},$ ${\bf v}$ and ${\bf B}$
 are all parallel.
 
 There is no dynamical friction in that case  because 
 deflection of a fluid element 
 towards the protoplanet requires motion perpendicular to the magnetic field lines.
 
 \noindent In contrast when $\psi =\pi/2$ deflection towards the protoplanet
 can occur with the motion  remaining parallel to the field lines.
 Then the interaction is uninhibited.
 
 \noindent Finally when $f_1 < 1$ and $f_2 =0,$ so that both waves exist but
 the ratio of propagation speeds is large, we obtain
 
 \noindent ${\cal D} = 1- f_1^2(\cos^2\psi -(\sin^2\psi)/2).$ In this case the most
 effective reduction of the dynamical friction occurs when $\psi =0,$ then ${\cal D} =1-f_1^2.$ 
  
   \subsection{General Scalings for the Response and Torque}

   \subsubsection{Conservation of Wave Action and Angular Momentum Evolution}\label{WA}
We point out that in the inviscid case considered here,  it is possible to derive 
a conservation law from equations (\ref{Lagp} - \ref{indp}) in the form 
\be {\partial \rho_A \over  \partial t} +\nabla\cdot ({\rho_A}{\bf v}+{\bf F_A}) = {\bf S_A} , \label{WVA}\ee
where in Cartesian coordinates with ${\bf r} = (x_1, x_2, x_3),$ $ {\bf v} = (v_1,v_2,v_3),$ 
$Im$ denoting that the imaginary part is to be taken and
using the summation convention
\be \rho_A = Im\left(\rho\boldxi^*\cdot{\partial \boldxi\over \partial t}
+\rho\Omega_* \boldxi^*\cdot({\bf {e_3}\times} \boldxi)
 -\boldxi\cdot{\partial  \left(\rho v_j\boldxi^*\right)\over \partial x_j}\right),
\ee
\be {\bf F_A} =  
Im\left({{\bf B'^*}{\bf \times (\boldxi\times B)}\over 4\pi}
 +{\nabla \times{\bf B}\cdot(\boldxi^*\times \boldxi){\bf B}\over 8\pi}
+ P'\boldxi^* \right)  \ee
and \be {\bf S_A} = -Im\left(\rho\boldxi^*\cdot \nabla \Phi'\right) .\ee
This conservation law is written for a general complex forcing potential
 Here we   apply it  to the situation
where the   forcing can be written as the  real part of a  potential $Re\left(\Phi'\right ),$ where 
$\Phi' =  g_m({\bf r})\exp (im(\omega_p t-\varphi)),$
$g_m$ is  a complex function of ${\bf r},$  $m$ the azimuthal mode number
and $\omega_p$ is the pattern speed, being the difference between the angular velocity
of the orbiting protoplanet and $\Omega_*.$
In this case, provided the forcing potential vanishes
at the fluid boundaries, the rate of increase of the energy of the fluid is
\be {dE_{fluid}\over dt}= \int Re(\rho')Re\left({\partial \Phi'\over \partial t}\right)d^3{\bf r} = {m\omega_p \over 2} \int {\bf S_A}d^3{\bf r}. \label{AEM}\ee
The quantities $   m\omega_p \rho_A/2 , m\omega_p {\bf F_A}/2$ may be regarded as the
energy density and the non advected   energy flux associated with the forcing.
In particular when the forcing source ${\bf S_A}$ is localised in space, the rate of change of orbital
energy is associated with a conserved energy flux propagating away to large distances from the protoplanet. The total effect of the protoplanet is obtained by summing the independent contributions from different $m.$ 

When  the unperturbed configuration is axisymmetric and the perturbing potential 
 $\Phi'$ depends on $\varphi$ and $t$  in the combination $ \omega_p t -\varphi,$
the energy fluxes convert to fluxes of the angular momentum 
component along the symmetry axis
by dividing by the pattern speed.

Then the total torque acting on the fluid can be written as 
${\cal T} = d J_{fluid}/ dt =\omega_p^{-1} dE_{fluid}/dt.$ 
When, as for the situations considered here, the wave fluxes  
produce an energy  and angular momentum loss from the system,
 the orbit decays or $dE_{orb}/dt = -dE_{fluid}/dt < 0.$  

We  comment that the fact that the energy
 changes in the orbit can be measured through energy fluxes
at distant boundaries generalises a corresponding result for hydrodynamic disk forcing
(see Papaloizou \& Terquem 2006) to the general MHD case.
In particular if there are no excited waves or advected disturbances, there are no induced
changes to the orbit. 
\subsubsection{Torque Scaling} 
    The scaling of the rate of energy change given by equation
    (\ref{bint}), that was obtained from the local analysis, 
    with the physical parameters of the problem can be obtained quite generally.
    The  length scale appropriate to the magnetosphere and the orbit  is
     expeced to be  $r_0.$ The unit of time is $\Omega_*^{-1},$ and the pattern speed is
     expected to be $\omega_p \sim \Omega_* ,$
     giving a characteristic relative velocity $r_0\Omega_*.$ 
      This velocity would be expected to be characteristic of both the flow velocity
      along field lines and the relative velocity between the orbiting protoplanet
      and the magnetosphere.   For a characteristic density  in the neighbourhood
      of the protoplanet, $\rho,$
    we find natural scalings from equations (\ref{Lagp} - \ref{indp}) and (\ref{AEM}) 
    for the response displacement and torque such that $\boldxi \propto r_0 M_p/M_*$ and 
     ${\cal T} \propto \rho (GM_p)^2/(\Omega_*^2 r_0).$
     Thus we  write 
     \be{\cal T} = {-1\over \omega_p} {dE_{orb}\over dt}=f\rho (GM_p)^2/
     (\Omega_*^2 r_0),\label{TN2}\ee
     where $f$ includes dependence on softening or the small scale cut off as well
     as the existence of propagating waves. This is equivalent to quation (\ref{bint})
     if
     $f  = 4\pi r_0\Omega_*^2 
     \ln\left ({k_{max}/ k_{min}}\right){\cal D}/(|{\bf v-U}|\omega_p )$

\section{Discussion}\label{sec5}
\subsection{Orbital Evolution Timescale}\label{Disc1}
From equation (\ref{TN2}) a characteristic rate of evolution of  protoplanet
 in circular orbit at radius $r_{orb}$ can be estimated from
\be t_{orb}^{-1} = \left |{d{E_{orb}/ dt}\over  E_{orb}}\right|
 \sim  { 2\rho G^2M_p r_{orb}\over \Omega_*^3r_0^4}\left(\left({r_{0}\over r_{orb}}\right)^{3/2} 
   - 1\right).\ee
For the magnetically dominated region (see eg. Bouvier et al. 2006)
we set  $\rho = {\dot M}/(4\pi r_0^3\Omega_*).$
Then for $M_* = 1M_{\odot}, M_p = 10^{-3}M_*, r_{orb} =0.5 r_0 $
giving a location slightly interior to the 2:1 commensurability with $\Omega_*,$ 
and  a characteristic accretion rate for a protostellar disk
 ${\dot M} = 3\times 10^{-8}M_{\odot} y^{-1}$
( e.g. Muzerolle et al. 2003 ), we obtain
$t_{orb} \sim 2\times 10^{11} y.$ 
Thus the expected evolution of such a protoplanet orbit is expected
to be very small over a characteristic protostellar disk lifetime
$\sim 10^{7} y.$  Note too that this result is not changed
if the orbital decay rate is enhanced by a factor $f\sim 50,$
corresponding to use of   equation (\ref{bint}) with ${\cal D} =1$
corresponding to effective wave propagation,
 $|{\bf v-U}| = (r_0\Omega_*)/2$
and $k_{max}/ k_{min}\sim 100.$
It also indicates that no additional mechanism, such as special torques
acting near the inner disk edge (Masset et al. 2006) is needed 
to halt the inward migration of proto giant planets.

  \subsubsection{Protoplanet Accretion}
  It may also be argued that the expected accretion  onto the
  protoplanet is  negligible while it is inside the magnetosphere.
  In order that accretion can take place   an amount of energy
  comparable to the orbital binding energy must be dissipated in 
  order for material to become bound to the protoplanet.
  Thus an estimate of the accretion  rate onto the protoplanet
  is given by
  \be {\dot M_p} 
 \sim  M_p \left|{1\over E_{orb}} {d E_{orb}\over dt}\right|,\ee
  which implies that the accretion time scale is the same as that for orbital
  evolution.

\subsubsection{Other Effects}
We here consider other effects that may lead to orbital decay
of the protoplanet. First we derive an effective drag coefficient
characterising  the dynamical friction acting through the gravitational torques
calculated above.

To do this we write
\be {dE_{orb}\over dt} =
 -{1\over 2}C_D\pi R_p^2 \rho |{\bf U} -{\bf v}|({\bf U} -{\bf v})\cdot {\bf U}
\ee
( Landau \& Lifshitz 1993), where the expression is applied locally
and  $C_D$ is the drag coefficient. f  Adopting $r_0/R_p = 200,$ we find  that the larger estimate of $t_{orb}$
given above corresponds to $C_D \sim 0.1$ while the smaller estimate corresponds to $C_D \sim 5.$
These results imply that the orbital decay rate is, to within an order of magnitude
or so, comparable to the non gravitational effects arising from the protoplanet
acting as an obstacle to the flow. This is in contrast to the
situation that occurs for protoplanets in circular orbits in a thin disk (Lin \& Papaloizou 1993).

\subsubsection{Magnetic Coupling with the Central Star}
Orbital energy decay
can also result from the direct interaction of a conducting
protoplanet with the stellar magnetic field when this varies around the orbit.
When the stellar magnetic field is non axisymmetric and the
protoplanet has a non zero resistivity , there will be periodic flux penetration
and dissipation leading to an orbital energy loss. In the discussion below
we neglect any internal protoplanetary magnetic field.
 
 The rate of change of energy  can be
estimated to be (see e.g. Joss, Katz \& Rappaport 1979 for a discussion of 
 magnetic binary systems)
\be {d E_{orb}\over dt} = - {3\over 8} R_p^2B_{||}^2 \Delta R_p (\omega_p -v_{\varphi}/r),\ee
where $\Delta R_P = \sqrt{\eta/(\omega_p -v_{\varphi}/r)}$
is a measure of the depth of external flux penetration per relative orbit,
with $v_{\varphi}$ being the azimuthal component 
of the gas velocity,  $B_{||}$ is a typical  component of ${\bf B}$ parallel to the
surface of the protoplanet, which we shall 
take to be $|{\bf B}|$  and $\eta$ is its magnetic diffusivity.
We may also  write

\be {dE_{orb}\over dt} =-{3\pi\over 2}\rho V_A^3R_p^2 
{ \Delta R_p(\omega_p-v_{\varphi}/r)\over V_A},\ee
which leads to an effective drag  coefficient  given by
\be C_D = {3V_A^2r(\omega_p-v_{\varphi}/r)
\over ||{\bf U} -{\bf v}|({\bf U} -{\bf v})\cdot {\bf U}|}
 \left({R_p\over r}\right)
\left({ \Delta R_p\over R_p}\right).\ee 

Although the flow is sub Alfv{\'e}nic,  the above  is of order unity apart from  the
 two factors in brackets. Of these $R_p/r \sim 10^{-2}$ and 
the last factor which represents  the ratio of the skin depth to radius
is also small. For example from
Zhang, Jones \& Chen (1996) the magnetic diffusivity of Jupiter can be estimated  to be
 $\eta = 3\times 10^3 cm^2 s^{-1}.$ For a relative orbital period of
$10 \ {\rm days},$ this leads to $ \Delta R_p / R_p \sim 10^{-5}$ and  the effective $C_D \sim 10^{-6}.$
Thus the effect of this type of 
magnetic interaction with the central star is not likely to be significant.

\subsection{Conclusions}

 In this paper we  performed 
a  local calculation of the protoplanet magnetosphere interaction derived.
The existence of propagating waves was necessary for there to be
 a non zero energy exchange rate.
 This is also expected from very general considerations of energy and
 wave action conservation. 
  For high
 relative speeds between the gas  and protoplanet, the interaction
 was found to be identical to that obtained from   the   dynamical friction
formalism of Chandrasekhar (1943).
But when, as expected here, the relative flow speed is below the fast magnetosonic speed but
  exceeds  the slow magnetosonic speed,  the interaction   strength was 
found to be reduced, especially when the relative velocity between gas and protoplanet 
was parallel or nearly parallel to the magnetic field.
The inhibition of the ability of the protoplanet to disturb the flow in such cases
would also be expected to apply to the gas accretion rate onto it.

For parameters
expected for protoplanetary disks our  calculation indicates
 negligible orbital evolution for protoplanets  interior
to magnetospheric cavities 
through which accretion is taking place during
expected protoplanetary disk lifetimes. This is the situation when the protoplanets
are far enough away from the inner disk edge so that interaction with the disk
is negligible and accordingly it is not necessary to invoke special torques associated 
with the disk edge to halt or reverse migration.
In this regard we comment that although 
there have been, of necessity, many simplifying assumptions
made in order to carry out the analysis presented here, the 
expected protoplanet orbital evolution in the magnetosphere 
 fails to be significant by a wide margin.



\label{lastpage}
\end{document}